\documentclass[sigconf,screen, authorversion]{acmart}

\copyrightyear{2023}
\acmYear{2023}
\setcopyright{rightsretained}
\acmConference[CUI '23]{ACM conference on Conversational User Interfaces}{July 19--21, 2023}{Eindhoven, Netherlands}
\acmBooktitle{ACM conference on Conversational User Interfaces (CUI '23), July 19--21, 2023, Eindhoven, Netherlands}\acmDOI{10.1145/3571884.3604319}
\acmISBN{979-8-4007-0014-9/23/07}

\begin{document}

\title{Why \emph{Are} Conversational Assistants Still Black Boxes? \\ The Case For Transparency}

\author{Trung Dong Huynh}
\email{dong.huynh@kcl.ac.uk}
\orcid{0000-0003-4937-2473}
\affiliation{%
  \institution{King's College London}
  \streetaddress{Bush House, 30 Aldwych}
  \city{London}
  \country{UK}
  \postcode{WC2B 4BG}
}

\author{William Seymour}
\email{william.1.seymour@kcl.ac.uk}
\orcid{0000-0002-0256-6740}
\affiliation{%
  \institution{King's College London}
  \streetaddress{Bush House, 30 Aldwych}
  \city{London}
  \country{UK}
  \postcode{WC2B 4BG}
}

\author{Luc Moreau}
\email{luc.moreau@kcl.ac.uk}
\orcid{0000-0002-3494-120X}
\affiliation{%
  \institution{King's College London}
  \streetaddress{Bush House, 30 Aldwych}
  \city{London}
  \country{UK}
  \postcode{WC2B 4BG}
}

\author{Jose Such}
\email{jose.such@kcl.ac.uk}
\orcid{0000-0002-6041-178X}
\affiliation{%
  \institution{King's College London}
  \streetaddress{Bush House, 30 Aldwych}
  \city{London}
  \country{UK}
  \postcode{WC2B 4BG}
}

\renewcommand{\shortauthors}{Huynh and Seymour, et al.}
\newcommand{\codeID}[1]{\texttt{\textbf{#1}}}
\newcommand{\tmplvar}[1]{\codeID{\color{red} #1}}

\begin{abstract}
Much has been written about privacy in the context of conversational and voice assistants. Yet, there have been remarkably few developments in terms of the actual privacy offered by these devices. But how much of this is due to the technical and design limitations of speech as an interaction modality? In this paper, we set out to reframe the discussion on why commercial conversational assistants do not offer meaningful privacy and transparency by demonstrating how they \emph{could}. By instrumenting the open-source voice assistant Mycroft to capture audit trails for data access, we demonstrate how such functionality could be integrated into big players in the sector like Alexa and Google Assistant. We show that this problem can be solved with existing technology and open standards and is thus fundamentally a business decision rather than a technical limitation.
\end{abstract}

\begin{CCSXML}
<ccs2012>
   <concept>
       <concept_id>10002978.10003029.10011703</concept_id>
       <concept_desc>Security and privacy~Usability in security and privacy</concept_desc>
       <concept_significance>500</concept_significance>
       </concept>
   <concept>
       <concept_id>10003120.10003121.10003124.10010870</concept_id>
       <concept_desc>Human-centered computing~Natural language interfaces</concept_desc>
       <concept_significance>500</concept_significance>
       </concept>
 </ccs2012>
\end{CCSXML}

\ccsdesc[500]{Security and privacy~Usability in security and privacy}
\ccsdesc[500]{Human-centered computing~Natural language interfaces}

\keywords{Conversational Assistants, Voice Assistants, Provenance, Audit Trails, Privacy, Transparency, Personal Data, Mycroft}

\maketitle

\section{Introduction}

Conversational assistants (CAs) promise everyday convenience in modern households through simple natural language commands. CAs work across text and voice (as voice assistants, or VAs), with vibrant ecosystems of first and third-party functionality (sometimes called `skills', `actions', and `capsules') growing around the latter. CAs can offer a vast range of voice-activated functionality, from checking emails to ordering a taxi, making online purchases, and even opening the garage door. Given the increased intermingling of text and voice-based conversational assistants (e.g. Google Assistant supports both modes of interaction) and the fact that voice commands are transcribed before being processed, we address both types of assistants in this paper. We use \emph{apps} to refer to first and third-party integration with conversational assistants.

The core appeal of conversational assistants, i.e.\ their use of language in lieu of a conventional graphical interface, is also a cause of confusion and concern: a common question from voice assistant users in the literature, for example, is that they don't~\cite{abdi2019more} know~\cite{10.1145/3170427.3188448} what~\cite{10.1145/3313831.3376529} their~\cite{10.1145/3274371} assistant~\cite{10.1145/2556288.2557421} shares.
This lack of transparency is (ironically) partly due to usability concerns; good conversational UX practices require that dialogue from a device be short, thus conveying much less information than would be via a GUI.\@ 
There is also an established norm that, with exceptions such as timers, assistants should be \emph{reactive}, speaking only when spoken to rather than proactively engaging the user~\cite{10.1145/3469595.3469629}.
 
At the same time, commands are often not authenticated for devices using voice; anyone within a voice assistant’s earshot can issue such commands, permitting unauthorised requests to be actioned without proper verification.
These devices also suffer from `phantom activations', where the assistant begins recording after hearing something similar to its wake word.
In this situation, other conversations in the vicinity can be captured and sent to virtually any app. 
Therefore, there can be cases where voice assistants disclose sensitive personal data to, for example, visitors, or they carry out instructions that adversely impact their owners’ finances or security.
Such incidents could happen without the owners' knowledge unless safeguards were established beforehand.
This is concerning because CAs frequently use personal data, much like smartphone apps do, with third-party integrations asking for permission often only on first use.

Privacy and transparency for these systems are, therefore, key concerns, but previous privacy-focused tools have often focused on visualising data usage, and matching embedded SDKs and network traces with real-world organisations (e.g.\ for smartphone apps~\cite{10.1145/3173574.3173967, 10.1145/3167996.3167999} and smart home devices~\cite{10.1145/3313831.3376264}).
Applying the same methodology to promote passive awareness of data sharing in conversational assistants, however, presents a number of problems due to the architecture of these devices. Firstly and most obviously, all data from an assistant is routed via the assistant's vendor before being sent onward to third-party services (this is simultaneously a security measure that ensures all communication from the device is encrypted, required for request transcription and intent-matching, and to mitigate the low specification hardware used in some standalone devices).
Secondly, as conversational apps are hosted \emph{by developers} and accessed by the device vendor through an API, there is no access to source code unless developers choose to open-source their voice apps.
This has previously been identified as a problem for platforms, who similarly have no way to check where data is sent once it has been passed to a third-party conversational app~\cite{10.1145/3543829.3544513}.
This leaves permissions requests and privacy policies as the only available information, which are difficult to read~\cite{10.1145/2470654.2481371} and may not accurately inform users about data sharing in practice (i.e.\ data collection and processing in the privacy policy often under- or over-specifies what the voice app actually does, as seen in~\cite{edu2022measuring}).

Against this background, we introduce a proof of concept that shows how actions taken by a VA can be routinely logged in a way amenable to effective auditing. Our implementation of \emph{data provenance} standards for the open-source voice assistant Mycroft allows for the discovery of undesirable data sharing by the VA and, ideally, the prevention of such instances through comparison with users' established norms. We then discuss how such technology could be used in commercial voice assistant platforms and lay out our vision for the voice assistants of the future.

\section{Proof of Concept: Provenance Tracking Inside Mycroft}

The real-world behaviour of a VA can only be analysed when they are observed with complete information about the chain of events that led to its actions.
A pizza order (for example) might, on the surface, appear to simply be the result of a spoken request.
However, under the hood the VA first needs to recognise the speech, interpret the user's intent, and choose the right app to execute it (potentially with auxiliary data to customise the request).
The invoked app may need to retrieve the user's stored preferences, ask the user for more information, or consult another third-party service to fulfil the request.
In this regard, provenance, traditionally used to establish the quality of food and the authenticity of art, can serve as the basis from which to capture a VA's actions and all the factors that influence it in scenarios like the one described above.
Paraphrasing the definition of provenance by the World Wide Web Consortium (WC3)~\cite{w3c-prov-dm}, we define the provenance of a VA action as ``a record that describes the people, institutions, entities, and activities involved in producing, influencing, or delivering'' that action.
It provides us with not only a record of an action taking place but also its \emph{history}, i.e.\ what happened that led to that action, which can be used to form assessments about it.
In other words, the provenance of a VA action is its audit trail, which can be represented as a knowledge graph linking the action back to its influences, be it some input data entities, a person or a piece of software.

\begin{figure}
\centering
\includegraphics[width=\columnwidth]{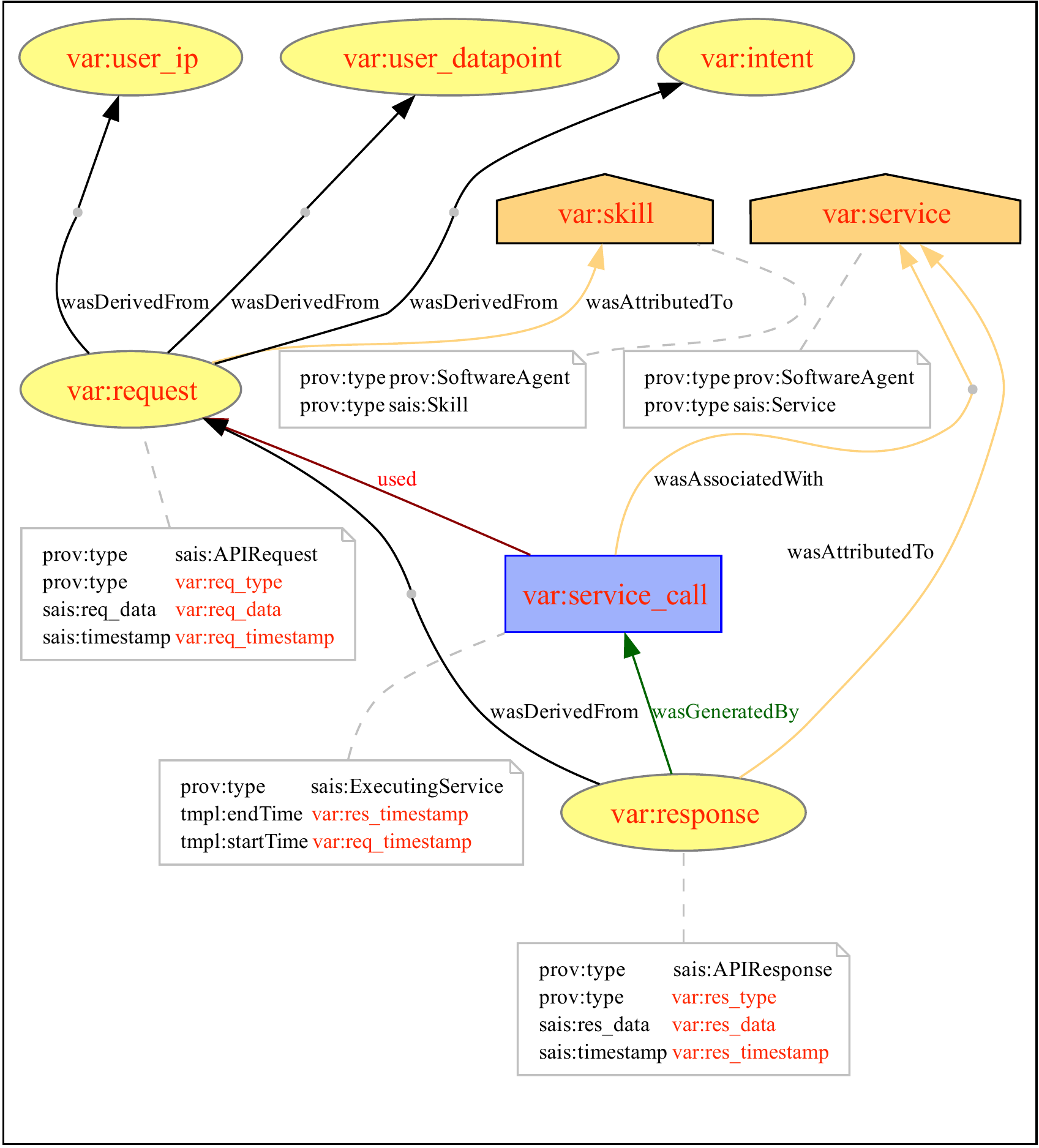}
\caption{The \texttt{skill\_invocation} provenance template, represented as a knowledge graph, for capturing a \tmplvar{request} with some \tmplvar{user\_datapoint} by a \tmplvar{skill} to an external web \tmplvar{service} in response to an \tmplvar{intent} of the user. The red texts with the \tmplvar{var:} prefix indicate variables to be replaced with runtime data to be recorded from an actual skill invocation.}\label{fig:template_si}
\end{figure}

To facilitate the systematic capturing of the provenance of VA actions, we adopt the PROV-Template approach~\cite{Moreau2018} to \emph{decouple} the modelling of the provenance (the expected shape of the audit trails) and its recording.
It allows us to describe the information to be recorded in a set of \emph{provenance templates} that contain provenance records with placeholders, called \emph{variables}, to be filled in later by values logged within a VA during its runtime, called \emph{bindings}.
Fig.~\ref{fig:template_si}, for example, presents a provenance template\footnote{The provenance template is shown as a (provenance) graph in Fig.~\ref{fig:template_si}. In such graphs, yellow ellipses represent \emph{entities}, e.g.\ input and output data points; brown pentagons represent \emph{agents}, e.g.\ a person or a software component; and blue boxes represent \emph{activities}, i.e.\ something that occurs over a period of time. They are linked to each other by provenance \emph{relations}, the labelled edges in the graph. In brief, entities are used and generated by activities, which are associated with agents. See~\cite{w3c-prov-primer} for a more detailed introduction to such provenance concepts.} describing an event in which some \tmplvar{skill} sends a \tmplvar{request} to an external \tmplvar{service} in response to a user's \tmplvar{intent}.
The red texts indicate \tmplvar{variables} to be replaced with data recorded from an actual skill invocation.
For instance, the identities of the actual skill (\tmplvar{skill}) and the request (\tmplvar{request}), i.e.\ the bindings for those variables, are to be logged at run time.
When the full provenance (of an action) is required, the recorded runtime data will be used to expand, or instantiate, the corresponding provenance templates to produce the  audit trail of a VA action.

In order to provide an example of what privacy and transparency tools could look like for voice assistants, and to back up our claims that their omission is a corporate choice, we instrumented Mycroft, an open-source VA, to log skill activations and the data flows within the VA, including the external services that receive the data. This involves constructing audit trails of logged skill activations in the form of provenance graphs, and generating explanations to answer user queries about historic skill activations, such as which services got the user's data. 
Note that voice apps in the Mycroft ecosystem are referred to as `skills' and, hence we use the latter when discussing voice apps in the context of Mycroft.

\subsection{Tracking Actions Inside Mycroft}\label{sec:provenance}

\begin{figure}
\centering
\includegraphics[width=\columnwidth]{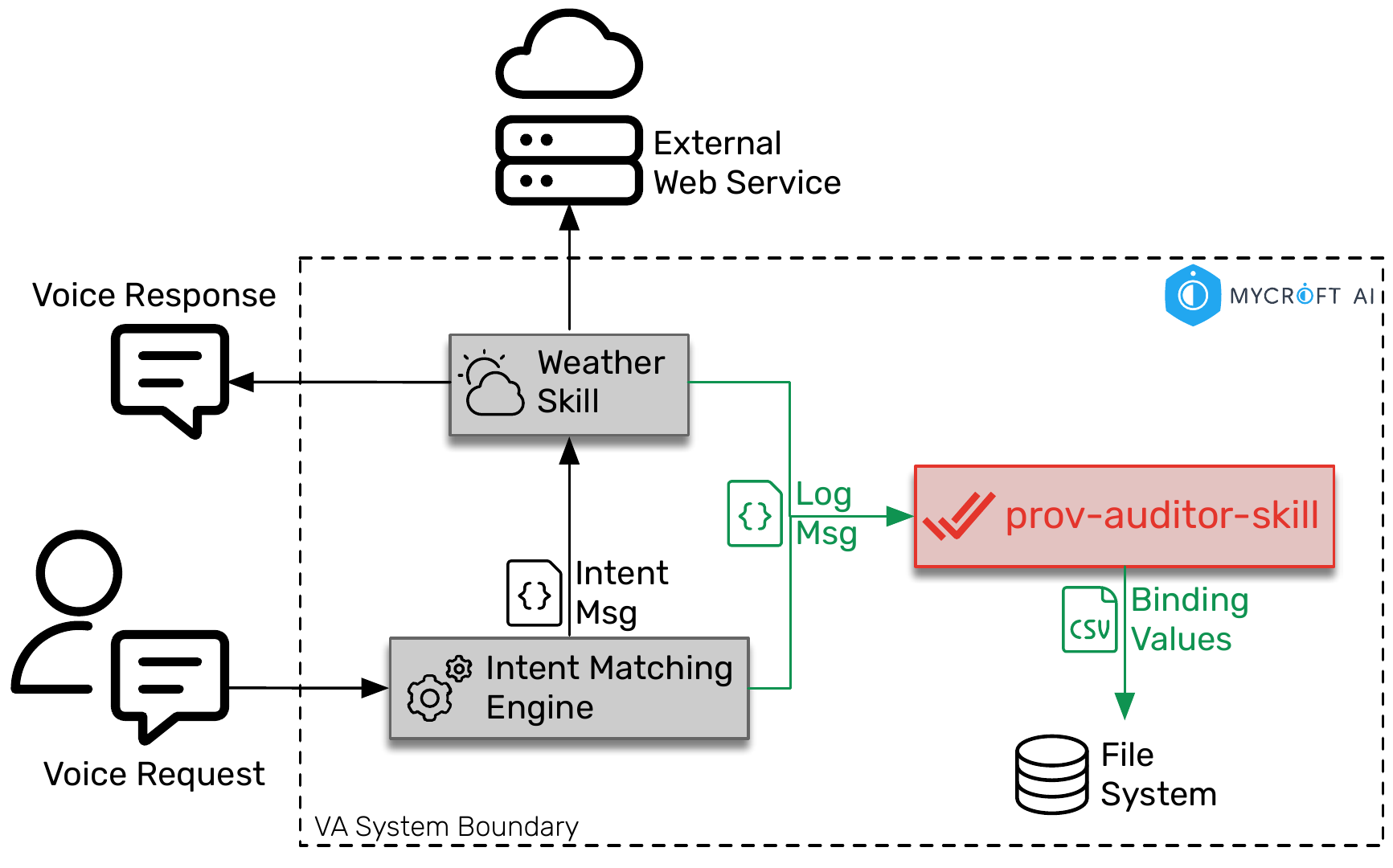}
\caption{Custom log messages capturing data from intent matching, skill invocation, and skill response are sent to \texttt{prov-auditor-skill} and transformed into stored bindings.}\label{fig:logging}
\end{figure}

\begin{figure*}
\centering
\includegraphics[width=.8\textwidth]{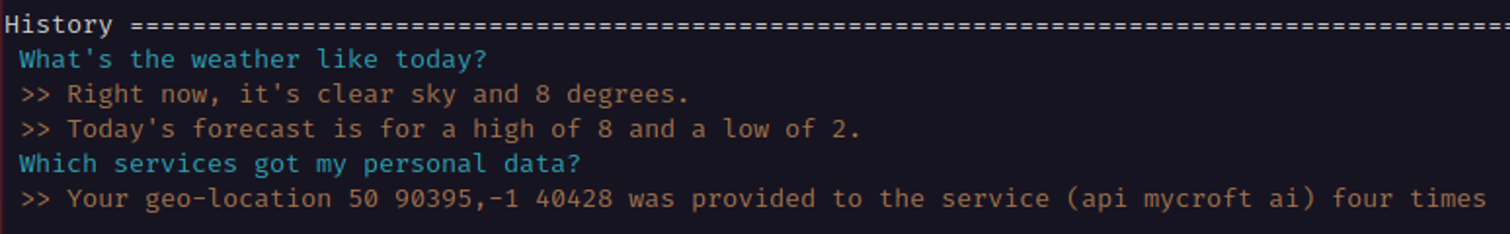}
\caption{Mycroft backend interface showing a user querying \texttt{prov-auditor-skill} about what happened with their data.}\label{fig:example}
\end{figure*}

We designed the following provenance templates to capture the provenance of skill actions within Mycroft:
\begin{itemize}
\item \texttt{skill\_invocation} template records the skill invoked by the inferred intent, the external web request it made, the user data it sent, and the response it received (see Fig.~\ref{fig:template_si}).
\item \texttt{sa\_response} template records the skill's action, if any, and the voice response played by the VA having executed the intent it received.
\item \texttt{intent\_matching} template records the user's intent inferred by the VA from a user's voice request.
\item \texttt{user\_datapoint} template describes specific data linked to the user, such as the data type and value.
\end{itemize}
To record the bindings for the templates, we created a Mycroft skill called \texttt{prov-auditor-skill}.
Once installed, it is loaded when a VA starts and continuously listens to custom Mycroft messages of specific types (see Fig.~\ref{fig:logging}).
Each of the custom message types was designed to capture the data needed to construct the bindings for each of the above templates (apart from the \texttt{user\_datapoint} template, whose bindings can be constructed directly from a Mycroft VA's configurations).
To produce such messages, instrumentation code was added to relevant points in Mycroft, such as its intent-matching procedure, and to the skills we want to track their actions (to observe the data it sends to an external service and the response it receives).
Data in those messages are converted by \texttt{prov-auditor-skill} into bindings for the corresponding templates and persisted onto the file system.
Over time, the user's requests and the responses they receive from specific skills are automatically logged in the form of bindings, ready for the construction of the audit trails for the skills' past responses (and actions) when needed.

\subsection{Querying Past Actions}\label{sec:explanations}

With skill actions and data flows inside a VA logged (in bindings), a user can query its past actions in various aspects.
For example, they could ask how often a skill was used, or where the VA sent their data.
The latter example is shown in Fig.~\ref{fig:example}, in which a user interacted with \texttt{prov-auditor-skill} asking ``Which services got my personal data''.
In order to respond to this question, \texttt{prov-auditor-skill} first constructs the audit trails from the recorded bindings, the result of which is a knowledge graph represented in the PROV data model~\cite{w3c-prov-dm}.
A graph query is run on the audit trails to find all data entities of type \texttt{sais:UserData} that was used to retrieve a \texttt{sais:APIResponse} entity.
The types \texttt{sais:UserData} and \texttt{sais:APIResponse} were previously specified in the \texttt{skill\_invocation} template (Fig.~\ref{fig:template_si}).
The voice response's narrative, ``Your geo-location \ldots'', is then produced by a natural language generation (NLG) engine from a language syntax tree crafted by us, taking into account the query's results, i.e.\ the actual data sent and the service that received it.
Additional queries and response narratives can be developed to provide users with more details from the recorded provenance following the methodology described in~\cite{huynh2022explainabilitybydesign}. 
Not only does the recorded provenance allow users to peek inside the black box of their VAs via supported queries, but it also serves as a foundation to further the accountability and safety of VAs, as discussed in the following section.

\section{Building Provenance Into Platforms}

An important question with any proof of concept is how such a system could be implemented at scale.
We chose Mycroft as the platform for the proof of concept because the project is open source, allowing us to modify the core of the assistant as well as the voice apps that run on it.
Additionally, Mycroft's architecture does not separate the assistant and its skills in the same way that larger platforms like Alexa and Google Assistant do, allowing data across all skills to be captured as required.
Turning now to how provenance could be implemented in large-scale commercial systems, we outline two levels of functionality---one simpler but easy to implement, and one more involved.

\subsection{Platform-level Provenance}

A good first step towards introducing provenance to conversational assistants would be to log access to personal data at the platform level. Developer policies for major platforms require that personal data is accessed through their permissions APIs whenever it is used (as opposed to being stored or cached by the application). Thus, VA platforms can easily log which intent was most recently invoked when an app requests user data and what data was supplied to the app. User questions around the use of personal data by the VA, provided by the \texttt{prov-auditor-skill} in our proof of concept, would instead be provided as a first-party app (i.e. this is not something that third parties can provide in the same way that might be possible for other platforms like web browsers that expose greater customisability). This would allow users to see which apps accessed their personal data, when that data was accessed, and which intent triggered the access. This is similar to the App Privacy Report available on iOS devices. However, it does not leverage the full potential of provenance to show where personal data is sent once it crosses the platform/app boundary. The only situation in which this approach would not work is where voice apps ask for personal information directly within a conversation rather than through a permissions API (though this is rarely seen in the wild as it contravenes the policies of major CAs)~\cite{edu2022measuring}.

\subsection{App-level Provenance}

In contrast, embedding provenance tracking into voice apps themselves would allow for tracking data as it moves into and out of apps, including to external destinations. Expecting developers to log data use for provenance tracking is a huge undertaking and unlikely to work across large app stores designed for size and heterogeneity. Instead, the most viable way to gather such information would be using static analysis by platforms with access to app source code (as is the case on smartphone apps with App Actions for Google Assistant). Code would be analysed upon submission to the app store, tracking the use of user data in external API calls in the same way that type checking is done for statically typed programming languages. These are then associated with a log event tied to the user's data and external resources.

Unfortunately, for CAs where source code is not visible to vendors and runs off-platform, there is no way to enforce the logging of provenance data (e.g. in Alexa skills). Providing provenance data would be a choice left to developers in these cases. However, the presence of alternatives shows that the decision to architect these platforms in this way is, in itself, a design choice made at the expense of privacy and transparency~\cite{10.1145/3543829.3544513}.

\subsection{What Could We Do With This?}

Logging provenance data allows for something that has thus far been a pipe dream in terms of platform transparency---allowing users to converse naturally with their assistant on the topic of data privacy. Previous explorations of data transparency document the shift in questioning as people develop an understanding of how their data is used, from \emph{who} receives data to \emph{what} data they receive and \emph{how} they use it~\cite{10.1145/3313831.3376264}. This information has not been available to previous data visualisation tools but is generated by provenance, particularly the app-level provenance described above. Prior work additionally suggests that the inclusion of privacy information in explanations about assistants is linked to lower privacy concerns, i.e.\ that in many cases these concerns are driven more by uncertainty than by specific disclosures~\cite{10.1145/3579497}.

This would facilitate asking about data usage whilst interacting with an app (via system-level intents), or even the provision of a privacy report delivered on demand or at regular intervals detailing data usage and associated justifications from assistant apps. This, in turn, provides an entry point for people to change their minds and modify/revoke permissions in response to knowledge about how apps actually use that data. 

\subsection{Towards Behavioural Norms}

Beyond increasing transparency around data usage, for what other purposes could provenance data be employed? One emerging area of study in this direction is the use of norms to evaluate the likelihood that any given use of data is genuine. Norms are commonly referred to when discussing the behaviour of devices and the people around them (e.g.\ in \citeauthor{nissenbaum2009privacy}'s theory of privacy as contextual integrity~\cite{nissenbaum2009privacy}), but capturing those norms can prove challenging. Recent work has explored the generation of norms around data use by voice assistants~\cite{10.1145/3411764.3445122}, and how these can be derived from user preferences~\cite{amoros2023predicting, 10.1145/3514094.3534129}.

Consider an example of a garage door controlled by natural language commands via a voice app within a VA. The assistant and the app log provenance data whenever they are invoked. Over time, norms like the following might be mined from the knowledge graph created from the recorded audit trails:
\begin{itemize}
    \item On weekdays the door is open for 2--5 minutes in the morning at around 08:00 and in the afternoon at around 18:00 (as the owner goes to and returns from work).
    \item Sometimes the door opens on weekdays for the same amount of time but at 05:00 or 06:00 (when the owner has to travel for work).
    \item During weekends, the door can be open for longer periods between 09:00 and 22:00 (e.g.\ for occasional car washing).
\end{itemize}
It may then be useful to notify users of the assistant if these norms are violated, such as the door being requested to open at night or at the expected time but staying open for much longer than usual.
    
\section{Conclusion}\label{sec:conclusion}

Given the multiple stakeholders controlling individual constituents of an assistant's ecosystem, producing complete and useful audit trails of its actions is a considerable technical and business challenge.
The prototype shown above, implemented for the open-source Mycroft VA, demonstrates that this is not an intractable problem with existing technology and open standards. However, reaping the full benefits of such audit trails requires enablers from multiple stakeholders for the presented vision of VA transparency, accountability, and safety to become a reality.
At a minimum, platform owners will need to provide app developers with the tools and incentives to report the provenance of data used in their applications.
They may also need to revise their architectures to allow for a more robust analysis of app code (and simultaneously address other privacy and security concerns).

An open question with the knowledge provided by audit trails, particularly for voice assistants, is how that data is accessed and used, and by whom. Provenance audit logs, by definition, contain personal data, and it is clear that this functionality should only be used with some means of authentication, such as voice profiles tailored to users or a smartphone-based prompt. Risks could be further mitigated by reducing the accuracy of data points (e.g.\ generalising locations or reporting times to the nearest half-hour). Finally, trust in the platforms behind conversational assistants plays an important role and is generally low after being undermined by a company's track records on privacy; even if CAs were capable of collecting the data required to provide audit trails that would strengthen personal privacy, it is not clear that they would be granted permission to do so.

\balance

\begin{acks}
This work is funded by the Secure AI Assistants project via Grant EP/T026723/1 from the UK Engineering and Physical Sciences Research Council (EPSRC).
\end{acks}

\bibliographystyle{ACM-Reference-Format}
\bibliography{main}

\end{document}